\documentclass{ifacconf}

\usepackage{natbib}        % required for bibliography

\usepackage{psfrag} 
\usepackage{longtable}
\usepackage{graphicx,verbatim,fancyhdr}
\usepackage{amssymb}
\usepackage{amsmath}
\usepackage{times}
\usepackage{color}

\begin{document}
\begin{frontmatter}

 \title{Architecture for communication with a fidelity criterion in unknown networks}

%\title{Style for IFAC Conferences \& Symposia: Use Title Case for
 %Paper Title\thanksref{footnoteinfo}} 

\thanks[footnoteinfo] {This work was supported by NSF Grant CCF-0836720, ÒCollaborative Research: CDI-Type
II: Discovery of Succinct Dynamical Relationships in Large Scale Biological Data SetsÓ, NSF Grant
ECCS-0801549, ÒControl over NetworksÓ, and Siemens Corporate Research Grant, ÒAdvanced Control
Methods for Complex Networked Systems.Ó}

%{Sponsor and financial support acknowledgment
%goes here. Paper titles should be written in uppercase and lowercase
%letters, not all uppercase.}

\author[First]{Mukul Agarwal} 
\author[Second]{Sanjoy Mitter} 

%\address[First]{National Institute of Standards and Technology, 
%   Boulder, CO 80305 USA (e-mail: author@ boulder.nist.gov).}
\address[First]{Laboratory for Information and Decision Systems, EECS, MIT, Cambridge MA 02139 (e-mail: magar@mit.edu)}
\address[Second]{Laboratory for Information and Decision Systems, EECS, MIT, Cambridge MA 02139 (e-mail: mitter@mit.edu)}

\begin{abstract}         
We prove that in order to communicate \emph{independent} sources between various users over an \emph{unknown} medium to within various distortion levels, it is sufficient to consider source-channel separation based architectures: architectures which first compress the sources to within the corresponding distortion levels followed by reliable communication over the unknown medium.
       % Abstract of not more than 250 words.
%These instructions give you guidelines for preparing papers for IFAC
%technical meetings. Please use this document as a template to prepare
%your manuscript. For submission guidelines, follow instructions on
%paper submission system as well as the event website.
\end{abstract}

\begin{keyword}
Source-channel separation, rate-distortion, digital communication, communication networks
\end{keyword}

\end{frontmatter}

\section{Introduction} \label{Introduction}

Architecture, defined as organization of distributed algorithms in software and hardware, plays a fundamental role in communications, control and computer science. The Von Neumann architecture of a stored program computer still today provides the model of computation. The separation theorem for source and channel coding in Shannon's theory of information provides an architecture for point-to-point communication.

Consider a controlled finite-state Markoff process $(X_t(u(\cdot)))_{t \geq 0}$ where $X_t$ is the state of the Markoff process at time $t$. The control at time $t$ is $u_t$. Let $Z_t$ represent a ``partial'' observation of the state at time $t$. It is required to choose the control function $u_t$ at time $t$ based on the past observation $(Z_s | 0 \leq s \leq t)$ in order to minimize the expected cost $J(u(\cdot)) =  E \int _0^T c(X_t, u_t) dt $. An important theorem states that the control separates into an estimation part, namely, computing the conditional distribution $\pi_t^u(X_t | Z_s, 0 \leq s \leq t)$ and then computing the optimal control $u^*_t$ by minimizing $J$ considered as a function of the information state $\pi_t^u$. Again, this leads to an architecture where the controller separates into an estimator and a controller. These are all examples of ``layered'' architectures. 

In this paper, we consider the question: how does one accomplish communication of various sources with a fidelity criterion, that is, to within particular distortion levels, over a common, unknown medium, optimally. This question arises in various contexts. A classic example is wireless: various users need to communicate via voice with each other over the unknown wireless medium and voice admits distortion. 

We answer this question under the following 3 assumptions:
\begin{itemize}
\item Distortion measures are additive
\item Sources that need to be communicated between various users are independent of each other. More precisely, for $(i,j) \neq (i', j')$, the source that needs to be communicated from user $i$ to user $j$ is independent of the source that needs to be communicated from user $i'$ to user $j'$. That is, \emph{the setting is unicast}
\item There is a shared source of randomness or common randomness at various users. Thus, random-coding is permitted
\end{itemize}

We prove that  digital communication is optimal to solve this problem. Digital communication is optimal is the same as saying that source-channel separation based architectures, that is, architectures where each user first compresses the source to within the corresponding distortion levels, followed by universal reliable communication of the resulting compressed source over the unknown medium, are optimal. \emph{There is optimality in the sense that if an architecture exists to accomplish this communication, a separation based architecture exists too.} Digital communication need not be optimal if there are other requirements (for example, some kind of robustness) in addition to the communication of the sources to within required distortion levels. 

The source-channel separation theorem that we prove is universal and holds for networks. \emph{Universality is over the medium of communication and not the source}. By universality, we mean that we do not need to know the exact operation of the medium: the medium is uknown. When modeled information theoretically, we mean that we do not know the precise operation of the network as a transition probability.

We do not provide any answers for the problem of reliable communication of bits over a network. This is the classical problem of network information theory. Our view is a reductionist view. \emph{We reduce the problem of rate-distortion communication over networks to the classical network information theory problem of reliable communication of bits  by showing the optimality of digital communication/source-channel separation architectures.}

Section \ref{PreviousWork} discusses the previous work on and related to this problem. Section \ref{ModelHigh} discusses the system model. The view that we will take to solve the problem described above  is discussed in Section \ref{Spirit}. This view is described behaviorally in Section \ref{Behavioral}. Section \ref{Distortion} defines various forms of communication. Section \ref{BasicTheorems} states and proves some theorems which will be helpful in proving our main result on universal source-channel separation for rate-distortion communication in networks in Section \ref{InformationTheory}. In Section \ref{Discussion}, we discuss our results with examples and conclude.

\section{Previous Work} \label{PreviousWork}

\cite{ShannonRD} proved that digital communication is optimal for  communication with a fidelity criterion in the point-to-point case. We differ in that we have solved the network version of the problem. Also, \cite{ShannonRD} does not solve the universal problem: the channel needs to be known. The universal point-to-point rate-distortion communication problem was solved by us in \cite{AgarwalSahaiMitter}. Furthermore, \cite{ShannonRD} requires some ergodicity assumptions on the channel whereas we do not require any ergodicity assumptions on the channel. We use a probability of excess distortion definition (\ref{PED})  for distortion over blocks compared to the expected distortion definition (\ref{EED}) used in \cite{ShannonRD}. This change of definition allows us to prove universal results for general, not necessarily ergodic channels.

In his thesis, \cite{Gastpar} proves optimality of separation architectures for certain networks, for example, when independent sources need to be communicated over a multiple access channel. Our work differs because we prove separation for general networks in the unicast setting, and not just in particular examples. We also prove separation in the universal context, unlike \cite{Gastpar}. Universality is possible, as was in contrast with \cite{ShannonRD}, because we use a different definition of  distortion over blocks. \cite{Gastpar} also contains examples when correlated sources need to be communicated over a network to within particular distortion levels. By two simple examples, it is shown in \cite{Gastpar} that separation architectures might not be optimal in this scenerio. The two examples are:
\begin{enumerate}
\item
Communication of correlated sources over a multiple-access channel
\item
Communication of the same source to within different distortion levels over a broadcast channel. In this scenerio, it is proved in \cite{Gastpar} that uncoded transmission can, in general, perform better than separation based communication. Note that the communication of the same source to two different users belongs to the multi-cast setting: the situation can be thought of as two different sources, which are infact identical, and hence, not independent, need to be communicated from a user to two other users.
\end{enumerate}
These examples show that in general, the unicast condition is necessary for separation architectures to be optimal. Our results and the results in \cite{Diggavi} which we discuss in brief below, show that the independence assumption is sufficient.

Tian, Chen, Diggavi and Shamai prove various results concerning optimality and approximate optimality of source-channel separation for rate-distortion in networks in \cite{Diggavi}. The result which has intersection with our result is where they prove optimality of separation based architectures when sources are independent of each other, over general networks. Results in \cite{Diggavi} are not universal. Results in \cite{Diggavi} require that the network have finite memory whereas we do not. As we stated above when comparing our work with \cite{ShannonRD}, these differences are made possible because we use a different definition of distortion.  \cite{Diggavi}  also contains interesting results for approximate optimality of separation architectures in the multi-cast setting: as the examples in \cite{Gastpar} show, in general, one cannot hope for optimality of separation architectures in this setting.

%Another difference between our view and the view in the above papers is that we have an operational view of source-channel separation for rate-distortion \cite{AgarwalSahaiMitterOperational}. The operational view is not discussed in this paper.

Separation is talked about in network-coding literature in the sense of separation of channel-coding and network-coding. See for example, \cite{KoetterMedard}. However, whenever me mention separation, we would mean the separation of source and channel coding.

%Network information theory is concerned with the problem  of reliable communication in networks. See, for example, the book  El Gamal and  Kim,  \cite{GamalKim}. A sub-branch of network information theory is network-coding where links are error free bit pipes limited to some capacity. See, for example, the book \cite{Yeung}. Separation is talked about in another terminology in the network-coding literature: the separation of channel-coding and network-coding. Consider a network which consists of noisy channels/links.  Channel/network coding separation holds for a link in a network if a  channel  of capacity $C$ in the network can be replace  with an error free link with throughput $C$ without changing the rate-region. For networks for which this happens, the network information theory problem of reliable communication can be reduced to the network-coding problem. \cite{KoetterMedard} contains an example of a situation where channel/network separation holds. Solution of reliable communication problems in networks is not our concern in this thesis. As stated before, our concern in this chapter is the reduction of rate-distortion communication problems in networks to reliable communication problem in networks, and we do this by proving the optimality of source-channel separation. Henceforth, when we mention separation, we would mean separation of source-coding  and channel-coding  and \emph{not} the separation of  channel-coding and network-coding. 

\section{Model of the given system} \label{ModelHigh}

\begin{figure}[!h]
\begin{center}
\includegraphics[scale = 0.6]{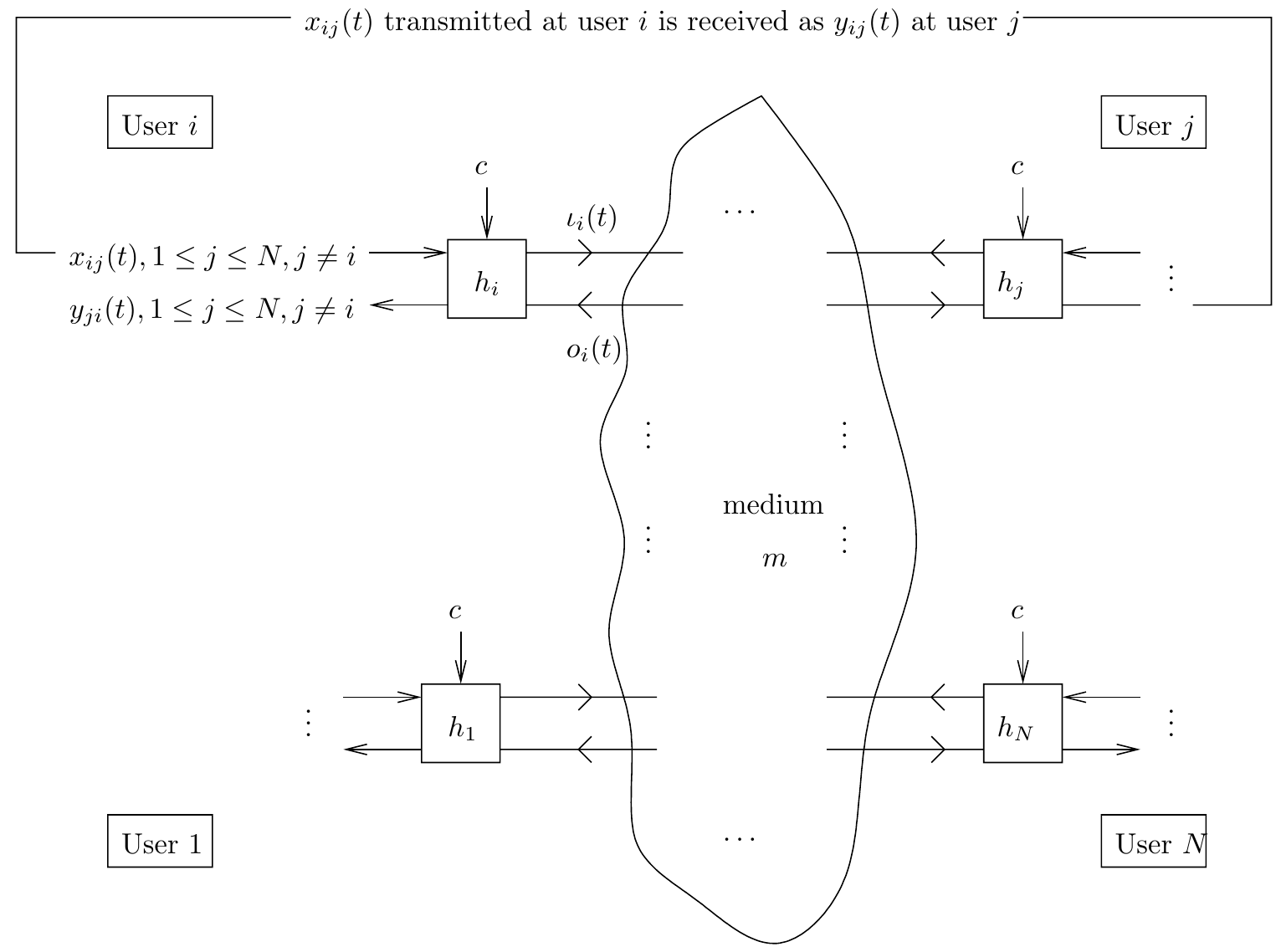}
\caption{System Model}
\label{SystemModelInterconnection}
\end{center}
\end{figure}

There are various users. The users communicate sources among each other. As shown in Figure \ref{SystemModelInterconnection}, the system consists of ``architecture boxes''  interconnected to a medium. The architecture boxes which will be refered to as modulators-demodulors or modems can be thought of as system protocol and aid communication.

More concretely:

There are $N$ users. $N$ might change with time. For $i \neq j$, user $i$ communicates source $X_{ij}(\cdot)$ to user $j$ over the system. The reproduction of $X_{ij}(\cdot)$ at user $j$  is $Y_{ij}(\cdot)$. $\forall t$, $X_{ij}(t) \in \mathcal X_{ij}(t)$ and $Y_{ij}(t) \in \mathcal Y_{ij}$. 

\emph{Note:} Above, when we mention $X_{ij}(\cdot)$, we mean the whole trajectory taken by the process over time $-\infty < t < \infty$. When we mention $X_{ij}(t)$, we mean the value at time $t$.

Note the ordering of $i$ and $j$ in $Y_{ij}(\cdot)$.

$m$ denotes the medium. $h_i, 1 \leq i \leq n$ is the modem at user $i$.

Modem $h_i$ at user $i$ takes source inputs $X_{i1}(\cdot)$, $X_{i2}(\cdot)$, $\ldots$, $X_{ij}(\cdot)$, $\ldots$, $X_{iN}(\cdot)$. $h_i$ takes input $I_i(\cdot)$ from the medium $m$. Modem $h_i$ produces an output $O_i(\cdot)$ into the medium $m$. In wireless systems, $I_i(\cdot)$ and $O_i(\cdot)$ are electromagnetic waves. Modem $h_i$  produces output  source reproductions  $Y_{1i}(\cdot)$, $Y_{2i}(\cdot)$, $\ldots$, $Y_{ji}(\cdot)$, $\ldots$, $Y_{Ni}(\cdot)$. $I_i(\cdot)$ is an input to the medium $m$ but output to the modem $h_i$. $O_i(\cdot)$ is an output of the medium $m$ but an input to the modem $h_i$. 

The modems are also assumed to have a common source of randomness denoted  by  $C$. The input $C$ is the same for all modems and can be used by the modems to generate random codes.

The medium takes inputs $I_1(\cdot)$, $I_2(\cdot)$, $\ldots$, $I_N(\cdot)$ and produces outputs $O_1(\cdot), O_2(\cdot), \ldots, O_N(\cdot)$.

The modem $h_i$ encodes information into input $I_i(\cdot)$. $I_i(\cdot)$ contains information about
\begin{enumerate}
\item
Sources $X_{ij}(\cdot), 1 \leq j \leq N$ that user $i$ wants to communicate to other users. 
\item
Sources $X_{i'j'}(\cdot), i' \neq i$. Modem $h_i$ has knowledge of other other sources $X_{i'j'}(\cdot)$ which are not inputs at user $i$ through the medium output $O_i(\cdot)$. \emph { In this case, information about $X_{i'j'}(\cdot)$ is being relayed through user $i$.   }
\end{enumerate}

Particular realizations of the random source processes  and their reproductions, and inputs and outputs to the medium will be denoted by $x_{ij}(\cdot), y_{ij}(\cdot), \iota_i(\cdot), o_i(\cdot)$. To avoid mathematical technicalities, \emph{it is assumed that the system evolves in discrete time, say, at every integer time}. For the same reason, it is assumed that the source alphabet, the source reproduction alphabet and the medium input and output alphabet is finite. 

Mathematically, the modem $h_i$ is a transition probability
\begin{align}
h_{i,\tau}
                        (y_{ji}(\tau), 1 \leq j \leq N,  i_i(\tau)
              \ |\  & \\ \nonumber
& \hspace{-1.5in}              
                                          x_{ij}(-\infty..\tau - 1), 1 \leq j \leq N, 
                                          o_i(-\infty^+..\tau - 1),  \\                        \nonumber
& \hspace{-1.5in}               
                                           c, y_{ji}(-\infty^+..\tau - 1), 1 \leq j \leq N,
                                           \iota_i(-\infty^+..\tau - 1))
\end{align}
denoting the probability that the
\begin{enumerate}
\item 
source reproduction output of modem $i$ at time $\tau$ are $y_{ji}(\tau), 1 \leq j \leq N$, 
\item
output produced by $h_i$ into the medium at time $\tau$ is is $\iota_i(\tau)$
\end{enumerate}
given 
\begin{enumerate}
\item
past source inputs are  $x_{ij}(t), -\infty < t  \leq  \tau - 1, 1 \leq j \leq N$,  
\item
past input from medium is $o_i(t), -\infty < t \leq \tau - 1$,
\item
common randomness input is $c$,
\item
past source reproduction outputs are $y_{ji}(t), 1 \leq j \leq N, 0 <  t \leq   \tau - 1 $, 
\item
past output into the medium is $\iota_i(t), 0 < t \leq  \tau - 1 $.
\end{enumerate}

Mathematically, the medium is a transition probability
\begin{align}
m_{\tau}
                        (  o_i (\tau), 1 \leq i \leq N
                            \ | \  
                            \iota_i(-\infty^+..\tau - 1), 1 \leq i \leq N,  \\ \nonumber
& \hspace{-1.5in}                            o_i(-\infty^+..\tau - 1), 1 \leq i \leq N, 
                            S) 
\end{align}
denoting the probability that the medium outputs at time $\tau$ are  $o_i(\tau), 1 \leq i \leq N$ 
given that 
\begin{enumerate}
\item
past inputs into the medium were $\iota_i(t), -\infty <  t \leq  \tau - 1, 1 \leq i \leq N$,
\item
past outputs produced by the medium were $o_i(t),  -\infty < t \leq  \tau - 1, 1 \leq i \leq N$,
\item
and that the initial medium state was $s$.
\end{enumerate}

The behavior of the medium $m$  may be complex. The interaction of medium $m$ and the modems $h_i$ and the resulting flow of information may be complex. The users may be co-operating. There may be multi-hopping and feedback. 

%The exact medium and system behavior does not concern us. In our formulation, the end-to-end system behavior that $X_{ij}$ is communicated from user $i$ to user $j$ and received as $Y_{ij}$ is what will matter. 

The sources $X_{ij}(\cdot)$ should be thought of as primitive in the sense that system behavior, that is, the behavior of the modems $h_i$ and the medium $m$ do not affect the sources.  This is a causality assumption.

The source waveform $x_{ij}(\cdot)$ is reproduced at a later time. $y_{ij}(t_m)$ is the reproduction of $x_{ij}(m)$ for some $t_m > m$. We define the process $y_{ij}[m]$ for integer $m$, denoted with square brackets by $y_{ij}[m] = y_{ij}(t_m)$. We also define $x_{ij}[m] = x_{ij}(m)$. In this notation, $y_{ij}[m]$ is the reproduction of $x_{ij}[m]$.

\section{Spirit of the question: High Level} \label{Spirit}

We ask a question in the following spirit.

\begin{figure*}
\begin{center}
\includegraphics[scale=0.7]{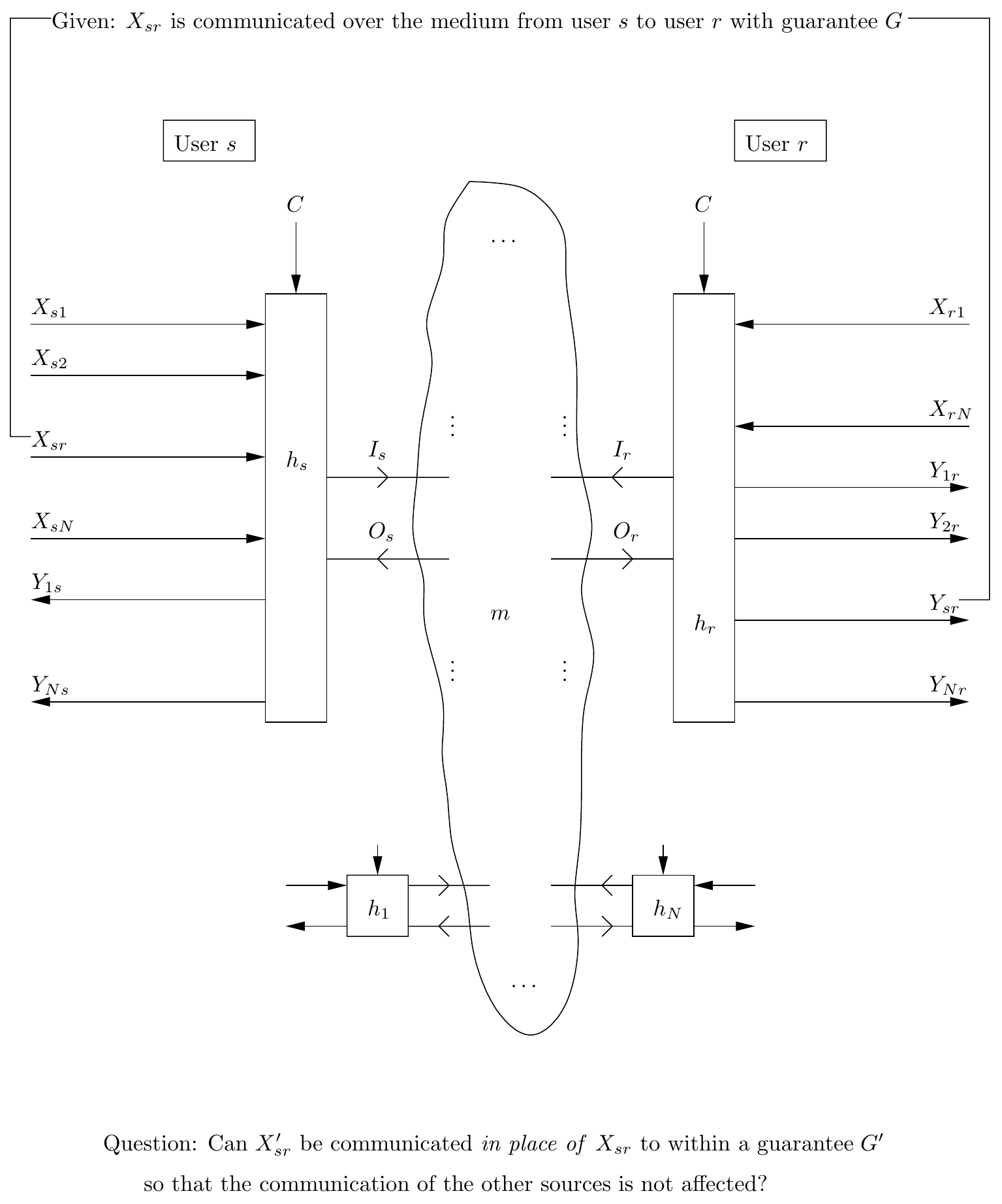}
\caption{}
\label{SpiritGiven}
\end{center}
\end{figure*}

Given a system as above. That is, a system which is known to communicate random sources $X_{ij}(\cdot)$ from user $i$ to user $j$, $1 \leq i,j \leq N$ over a medium. See Figure \ref{SpiritGiven}.

Let $s$ and $r$ be two particular users. It is known that source $X_{sr}(\cdot)$ is communicated from user $s$ to user $r$ over the system with some guarantee. Denote the guarantee by  $G$. $X_{sr}(\cdot)$ is received as $Y_{sr}(\cdot)$. An example of a guarantee and the one we will use is that $X_{sr}(\cdot)$ is communicated to within some distortion level.

We  ask a question about the communication of another random source $X'_{sr}(\cdot)$ evolving in time \emph{ in place of}  source $X_{sr}(\cdot)$ from user $s$ to user $r$. The source $X'_{sr}(\cdot)$ should be received with some guarantee $G'$ depending on $G$. The gaurantee $G'$ that we will use would be that $X'_{sr}(\cdot)$ needs to be communicated to the destination to within some distortion level.

\emph{We will assume that the sources $X_{ij}(\cdot)$ are independent of each other $\forall i,j$. This assumption is crucial.}

\emph{We will also assume that the source $X'_{sr}(\cdot)$ is independent of sources $X_{ij}(\cdot) \forall i,j$. }  In order to prove the result concerning optimality of digital communication as stated in Section \ref{Introduction}, it is okay to make this assumption. $X'_{sr}(\cdot)$ is primitive in the sense discussed in Section \ref{ModelHigh}.

The changes made in the system for the desired communication of $X'_{sr}(\cdot)$ from user $s$ to user $r$ should not change the communication of $X_{ij}(\cdot)$ from user $i$ to user $j$ for $(i,j) \neq (s,r)$. Mathematically, this means that $X_{ij}(\cdot)$ should be received precisely as  $Y_{ij}(\cdot)$ \emph{in distribution} for $(i,j) \neq (s,r)$. Of course, instead of $X_{sr}(\cdot)$, $X'_{sr}(\cdot)$ now, needs to be communicated from user $s$  to user $r$. \emph{$X_{sr}(\cdot)$ does not need to be communicated any more}.

Each user only has local knowledge. At time $\tau$, user $i$ has knowledge of the source realization  $x_{ij}(t), -\infty \leq t \leq \tau - 1 , 1 \leq j \leq N$, the modem $h_i$, medium input realization $\iota_i(t), -\infty < t  \leq \tau - 1$, medium output  realization $o_i(t), -\infty < t \leq  \tau - 1$, the realization of reproduction of sources from various users destined for user $i$, $y_{ji}(t), -\infty < t \leq  \tau - 1, 1 \leq j \leq N$ and the common randomness input $C$. User $i$ also has knowledge of any guarantees associated with sources at user $i$, that is, sources $X_{ij}(\cdot), 1 \leq j \leq N$. It is known to all users that sources $X_{ij}, 1 \leq i,j \leq N, X'_{sr}$ are all independent of each other.

 Users do not have knowledge of the medium kernel $m_\tau$ defined in the previous section.
 
System architecture can be changed, only locally. That is, $h_s$ and $h_r$ can be changed in order to communicate the source $X'_{sr}(\cdot)$. All other modems should remain the same. That is, for $i \neq s,r$, $h_i$ should remain unchanged.

Question: when can $X'_{sr}(\cdot)$ be communicated to with the required guarantee $G'$ and how.

The definitions of guarantees $G$ and $G'$ will be given later.

The communication of $X'_{sr}(\cdot)$ will be accomplished in the following way:

Since there is no knowledge of the medium kernel $m_\tau$, we would like to mantain the input-output behavior of the medium. If the joint input distribution of the medium inputs $I_i(\cdot), 1 \leq i \leq N$ is changed, in the absence of the knowledge of medium kernel, it is impossible to know the evolution of the medium outputs.  In order to mantain the medium joint input distribution, we would  mantain the distribution $X_{sr}(\cdot)$. We would build an encoder $e$ which would map the source $X'_{sr}(\cdot)$ into an encoded input whose distribution is precisely the same as the source process $X_{sr}(\cdot)$. We will thus simulate $X_{sr}(\cdot)$.  Denote this simulated source by $X^s_{sr}(\cdot)$. The guarantee $G$ will be satisfied between the simulated source $X^s_{sr}(\cdot)$ and output which we denote by $Y^s_{sr}(\cdot)$. We will then use this output $Y^s_{sr}(\cdot)$ to make a decoding $Y'_{sr}(\cdot)$ with the use of a decoder $d$. 

\emph{This encoding procedure  can be thought of as embedding information about $X'_{sr}(\cdot)$ into $X_{sr}(\cdot)$. }

\begin{figure*}
\begin{center}
\includegraphics[scale = 0.7]{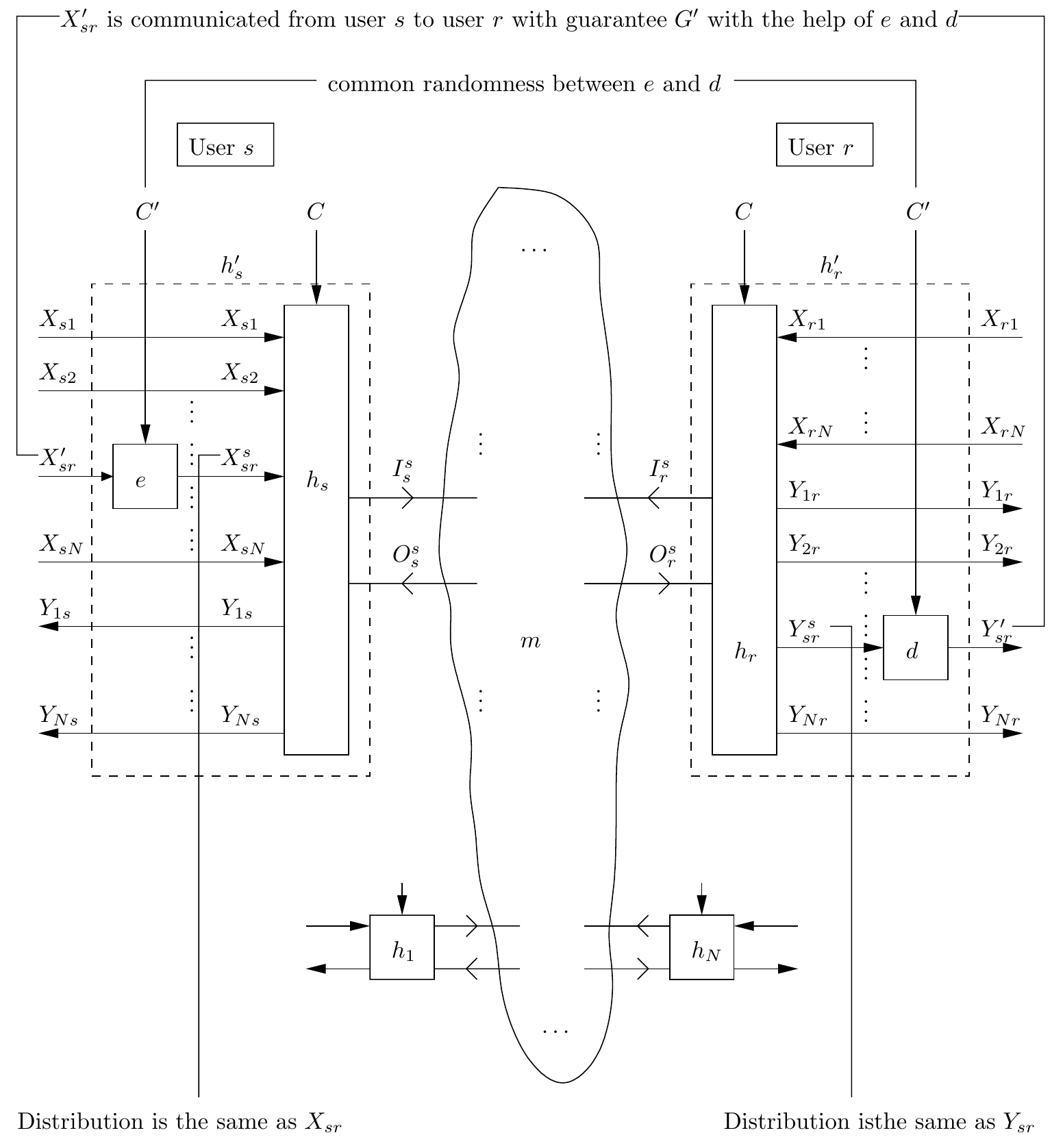}
\caption{}
\label{SpiritToAchieve}
\end{center}
\end{figure*}

Note that with this encoding-decoding procedure, we will not be ``disconnecting'' the modems $h_s$ and $h_r$ from the medium. The new modem $h'_s$ at user  $s$ is the composition of $h_s$ and $e$. The new modem $h'_r$ at user $r$ is the composition of $d$ and $h_r$. In other words, we are building ``on top of'' the existing architecture to accomplish the required communication. Note that $e$ is a stochastic code. As we shall see later, the way we will build the encoder-decoder $e-d$,  there would be need for  common randomness $C'$ between $e$ and $d$. That is, $e-d$ is a random code. See Figure \ref{SpiritToAchieve}.

By requirement, the modem $h'_i$ is the same as $h_i$ for $i \neq s,r$.

The joint distribution of the inputs to modems $h_i$ has been mantained. This is because $X_{ij}(\cdot)$ is unchanged for $(i,j) \neq (s,r)$. For $(i,j) = (s,r)$, the input, now is  $X^s_{sr}(\cdot)$ instead of $X_{sr}(\cdot)$. $X^s_{sr}(\cdot)$ has  the same distribution as $X_{sr}(\cdot)$. $X^s_{sr}(\cdot)$ is independent of $X_{ij}(\cdot), (i,j) \neq (s,r)$ by construction and because of the assumption that $X'_{sr}(\cdot)$ is independent of $X_{ij}(\cdot)$. Thus, the joint distribution at the inputs to modems $h_i$ has been mantained. As a result, $X_{ij}(\cdot)$ is received precisely as $Y_{ij}(\cdot)$ for $(i,j) \neq (s,r)$. $X_{sr}(\cdot)$ is not transmitted anymore, however. Instead, $X^s_{sr}(\cdot)$ is transmitted. 

We stated before that we would like the joint medium input and output distributions to be mantained. By mantaining the distribution of $X_{sr}(\cdot)$, this has automatically happened. 

%The medium input at user $s$ is $I^s_s$. $I^s_s$ has the same distribution as $I_s$. The medium output at user $s$ is $O^s_s$. $O^s_s$ has the same distribution as $O_s$. The medium input at user $r$ is $I^s_r$. $I^s_r$ has the same distribution as $I_r$. The medium output at user $r$ is $O^s_r$. $O^s_r$ has the same distribution as $O_s$. 

\emph{Note:} we are using this way of simulating $X_{sr}(\cdot)$ and ``building on top'' of the already existing architecture in order to communicate $X'_{sr}(\cdot)$ from user $s$ to user $r$. Other ways may exist. This is the view and method that we use.

The assumption of independence of sources $X_{ij}(\cdot)$ is required in the above construction for the following reason:

Let $X_{ij}(\cdot)$ and $X_{sr}(\cdot)$, $(i,j) \neq (s,r)$ be dependent. In order to communicate $X'_{sr}(\cdot)$, we simulate $X_{sr}(\cdot)$ as described above. This would mean that $X_{ij}(\cdot)$ would also need to be, atleast partially simulated in order to respect the joint distribution of $X_{sr}(\cdot)$ and $X_{ij}(\cdot)$. This would mean that the system behavior would change for the transmission of $X_{ij}(\cdot)$ from user $i$ to user $j$. This is not permitted.

Modem $h'_s$ consumes the same energy as the modem $h_s$. This is because the new medium input has the same distribution as $I_s(\cdot)$. We are neglecting any energy consumption in the circuits of the modem.  Also, the bandwidth of the medium consumed by the modem $h'_s$ is the same as the bandwidth consumed by the modem $h_s$. This is because the new medium input has the same distribution as $I_s(\cdot)$.

In general, consumption of all resources related to the medium remains unchanged if we mantain the marginal of $I_s(\cdot)$.

A similar procedure can potentially be followed for communication of other sources $X'_{ij}(\cdot)$ from a user $i$ to user $j$, $1 \leq i, j \leq N$ . This results in a decentralized system for communication of various sources between various users over a network. 

We will elaborate on, and see an application of the reasoning described in this section to prove a source-channel separation for rate-distortion in networks by making the source $X'_{sr}(\cdot)$ have the same distribution as the source $X_{sr}(\cdot)$. This section just describes the view.

\section{Behavioral view} \label{Behavioral}

In this section, we put the ideas discussed in the previous section in a behavioral perspective of Willems \cite{Willems}

By convention, a random variable $S$ taking values in a set $\mathcal S$ has a probability distribution denoted by $p_S$. 

\emph{Behavior of a stochastic system}: The behavior of a stochastic system $s$, $\mathcal B_s \subset \{ S \ : \ S \ \mbox{is a random variable taking values in } \ \mathcal S\}$. If, for example $\mathcal S = \mathcal R$, $\mathcal B_s$  is a subset of all random variables on  $\mathcal R$. If, for example, $\mathcal S = \mathcal R ^{[0,\infty)}$, $\mathcal B_s$ is a subset of all stochastic processes on $\mathcal R ^{[0,\infty)}$.

\begin{figure}[!h]
\begin{center}
\includegraphics[scale = 1.0]{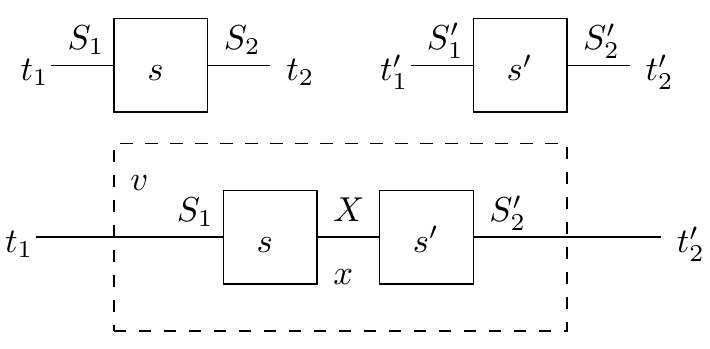}
\caption{}
\label{Interconnect}
\end{center}
\end{figure}

\emph{Interconnection of stochastic systems:} Let $s$ be a stochastic system with two ``terminals'' $t_1$ and $t_2$. The random variable at terminal $t_1$ is $S_1$, taking values in set $\mathcal S_1$.  The random variable at terminal $t_2$ is $S_2$, taking values in set $\mathcal S_2$. $\mathcal B_s \subset \{ S_1S_2 \ : \ S_1 S_2 \  \mbox{is a random variable taking values in } \ \mathcal S_1 \times \mathcal S_2 \}$. Similarly, let $s'$ be a stochastic system with two ``terminals'' $t'_1$ and $t'_2$. The random variable at terminal $t'_1$ is $S'_1$, taking values in set $\mathcal S'_1$.  The random variable at terminal $t'_2$ is $S'_2$, taking values in set $\mathcal S'_2$. $\mathcal B_{s'} \subset \{ S'_1S'_2 \ : \ S'_1 S'_2 \  \mbox{is a random variable taking values in } \ \mathcal S'_1 \times \mathcal S'_2 \}$. Let $\mathcal S'_1 = \mathcal S_2$. The interconnection of systems $s$ and $s'$, denoted by $v$, when terminal $t_2$ is connected to terminal $t'_1$ is defined behaviorally as follows: 
$\mathcal B_v = \{ S_1XS'_2 \ : \ S_1X \in \mathcal B_{s} \ \mbox{and} \ XS'_2 \in \mathcal B_{s'} \}$, where $X$ is the random variable at the terminal $x$ which is the interconnection of terminals $t_2$ and $t'_1$. See Figure \ref{Interconnect}

\emph{Primitive and non-primitive random variables:}  Primitive random variables are those which evolve autonomously. An example of a  primitive random variable is a source which needs to be communicated to a destination. Non-primitive random-variables come out of action of systems on primitive random variables. An example  of a non-primitive random variable is a source-reproduction.

Interconnection of stochastic systems, as defined above might \emph{not make physical sense} in certain cases.  

For example, consider the case when $S_2$ and $S'_1$ are independent primitive random variables. The above interconnection forces $S_2 = S'_1$. Even if $S'_1$ and $S_2$ had the same distribution, this interconnection does  not make \emph{physical} sense because $S_2$ and $S'_1$ are primitive and might not be or evolve in a way that they are equal to each other.  Such an interconnection might make sense if $S'_1$ were not primitive, for example, if $S'_1$ were an output of the system and equal to $S_2$. 

Consider another example when $S_1$ and $S'_2$ are primitive and independent. The above interconnection might cause  a dependence between the realization of $S_1$ and $S'_2$ which might not be consistent with them being independent. However, if the behavior of the system $s'$ were $\mathcal B_{s'} =  \{ S'_1S'_2 \ : \ S'_1\ \mbox{and} \ S'_2 \  \mbox{are independent} \}$, then, the above interconnection won't lead to inconsistency.

The previous section can be summarized in the behavioral view as follows. Systems $e$ and $d$ need to be constructed. System $e$ needs to be interconnected to $h_s$ and system $d$ needs to be interconnected to $h_r$ as shown in Figure \ref{SpiritToAchieve}. The following should be satisfied
\begin{enumerate}
\item
The process $X'_{sr}(\cdot)X_{sr}^s(\cdot) \in B_e$ such that  $X^s_{sr}(\cdot)$ has the same distribution as  $X_{sr}(\cdot)$
\item
$Y^s_{sr} Y'_{sr}  \in \mathcal B_d$ where  $Y^s_{sr}(\cdot), t \in (-\infty, \infty)$ has same distribution as $Y_{sr}(\cdot)$, and such that that the guarantee $G'$ is satisfied between the processes $X'_{sr}(\cdot))$ and $Y'_{sr}(\cdot)$. 
\end{enumerate}
Note that $X_{sr}(\cdot)$ is not primitive  any more because it is no longer a source: As stated in Section \ref{Spirit}, $X'_{sr}(\cdot)$ needs to be communicated \emph{in place of} $X_{sr}(\cdot)$. Otherwise, when interconnecting $e$ and $h_s$, we would have landed in the problem of the first example described above. Also, the reason for the assumption of the independence of sources $X_{ij}(\cdot), 1 \leq i,j \leq N$ made in the previous section is for precisely the same reason as the second example above. 

\section{Communication to within a distortion level and error} \label{Distortion}

Let $\displaystyle{d: \mathcal X_{sr} \times \mathcal Y_{sr} \rightarrow [0,\infty)}$ be a function. $d$ is the distortion function. For $x_{sr} \in \mathcal X_{sr}$, $y_{sr} \in \mathcal Y_{sr}$,  $d(x_{sr}, y_{sr})$ is the distortion incurred if $x_{sr}$ is decoded as $y_{sr}$.

\emph{Notation:}  $n$ length sequences will be denoted with superscript $n$. 

\emph{Definition:} 
Distortion between $n$ length sequences $x_{sr}^n \in \mathcal X_{sr}^n$, $y_{sr}^n \in \mathcal Y_{sr}^n$ is additive:
$\displaystyle{d^n(x_{sr}^n, y_{sr}^n) \triangleq \sum_{k=1}^n d(x_{sr}^n[k], y_{sr}^n[k])}$. Average distortion is $\displaystyle {\frac{1}{n}d^n(x_{sr}^n, y_{sr}^n) = \frac{1}{n} \sum_{k=1}^n d(x_{sr}^n[k], y_{sr}^n[k])} $.

\emph{Definition:} Let the source $X_{sr}(\cdot)$ be discrete and evolve  as $X^n_{sr}(1)$, $X^n_{sr}(2), \ldots$,  $X^n_{sr}(n)$. $n$ is the block-length. The source will be  denoted by $X_{sr}^n$ to make explicit the fact that the block-length is $n$. Recall the last paragraph of Section \ref{ModelHigh} that the source is also denoted as $X^n_{sr}[1], \ldots, X^n_{sr}[n]$. The reproduction of $X^n_{sr}[i]$ is $Y^n_{sr}[i]$. Source $X^n_{sr}$ of block-length $n$ is said to be communicated to within a distortion level $D$ under metric $d$ with error probability  $< \epsilon$ if 
\begin{align}\label{PED}
\Pr \left ( \frac{1}{n} d^n(X^n_{sr}[ \ ] , Y^n_{sr} [\ ]) > D \right ) < \epsilon
\end{align}
Probability is taken with respect to the joint distribution of $X^n_{sr}[ \ ]$ and $Y^n_{sr}[ \ ]$ which can be obtained by marginalization from the joint distribution $(X_{ij}, I_i, O_i, Y_{ij}, 1 \leq i,j \leq N, 0 < t < \infty)$.

\emph{Notation and definitions:} The source $X'_{sr}$ takes values in the set $\mathcal X'_{sr}$. The reproduction of $X'_{sr}$ is $Y'_{sr}$. $Y'_{sr}$ belongs  to the set $\mathcal Y'_{sr}$. Analogously as above, then, we can define a distortion metric $d'$ and talk about the communication of block-length 
$n'$ source $X'^{n'}_{sr}$ to within a distortion $D'$ under metric $d'$ with error probability $< \epsilon'$.

\emph{Notation and definition:} Let $\mathcal M^n$ be a message set of cardinality $2^{nR}$ for some $R$. The message $M^n$ is a random variable which has some distribution on $\mathcal M^n$. Note that $M^n$ does not necessarily have the uniform distribution. For our purpose, the precise distribution of  $M^n$ will not affect the results. We will ask a question about communication of $M^n$ from user $s$ to user $r$. Let $\hat M ^n$ be some decoding of $M^n$ after transmission over some system.  Rate $R$ source $M^n$ of block-length $n$ is said to be communicated with error probability $< \delta$ under the MBP criterion if 
\begin{align}
\sup_{m^n \in \mathcal M^n} \Pr(\hat M^n \neq M^n | M^n = m^n) < \delta
\end{align} 
MBP stands for Maximal Block Error Probability.  In the limit as $n \to \infty$, if $\delta \to 0$, we say that there is reliable communication at rate $R$ under the maximal block error probability (MBP) criterion.

\emph{Notation: In what follows, we will sometimes denote the process  $X_{ij}(\cdot)$ by just $X_{ij}$ and similarly for other processes}

\section{Basic Theorems} \label{BasicTheorems}

In this section, we will prove results concerning communication from user $s$ to user $r$. Communication does takes place between other users $i$ and $j$, $(i,j) \neq (s,r)$. We will not be concerned with communication beween $(i,j) \neq (s,r)$ in the sense that we do not want that communication to be affected by any changes that we make to the system for communication from user $s$ to user $r$. That is, even if we make changes to the system architecture, $X_{ij}$ should still be received precisely as $Y_{ij}$ if $(i,j) \neq (s,r)$. This was discussed in Section \ref{Spirit}. 

Recall the assumptions made in Section \ref{Spirit} that  the sources $X_{ij}(\cdot), 1 \leq i,j \leq N, X'_{sr}(\cdot)$ are independent of each other. We also assume that the random sources $X_{sr}(\cdot)$ and $X'_{sr}(\cdot)$ are i.i.d. The results can be generalized to stationary ergodic sources under some conditions. 

\emph{Notation:} Since we will be concerned only with communication between user $s$ and user $r$, in order to simplify notation, $\mathcal X_{sr}$, $\mathcal Y_{sr}$, $X_{sr}$, $Y_{sr}$, $X'_{sr}$, $Y'_{sr}$, $X^s_{sr}$, and $Y^s_{sr}$ will be denoted by $\mathcal X$, $\mathcal Y$, $X$, $Y$, $X'$, $Y'$, $X^s$, and $Y^s$ respectively.  

\emph{ Consider the source $X_{sr}^n$ of block-length $n$ which is denoted in simplified notation by  $X^n$. It is known that $X^n$ is communicated to within a distortion level $D$ under metric $d$ with error probability $< \epsilon$. from user $s$ to user $r$. This communication is the guarantee $G$ in the language of Section \ref{Spirit}. Consider the source $X'^{n'}_{sr}$ of block-length $n'$ which  is denoted in simplified notation by $X'^{n'}$.  We ask the question: can source $X'^{n'}$ of block-length $n'$ be communicated to within a distortion level $D'$ under the metric $d'$ with error probability $< \epsilon'_{n}$ from user $s$ to user $r$ in place of $X^n$ in the way described in Section \ref{Spirit}. This communication of the source $X'^{n'}$ is the guarantee $G'$ in the language of Section \ref{Spirit}.}

\emph{Note: We are operating in the framework of information theory. In particular, delays do not matter. Any decoding that needs to be performed can be performed  after observing the whole output process. That is, the decoding need not be causal.}
 
In order to answer this question, we first answer the question of communication of rate $R$ source $M^n$  defined in the previous section over the system under the MBP criterion with error probability $< \delta_n$. Another way of saying this in  the language of Section \ref{Spirit} is the following: $X_{sr} = X^n$. $G$ is the communication of source $X^n$ to within a distortion level $D$ under metric $d$ with error probability  $< \epsilon$.  $X'_{sr} =$ rate $R$ source  $M^n$. $G'$ is the communication of rate $R$ source $M^n$ under the MBP criterion with error probability $< \delta_n$.

\emph{Notation:} Let $R_X(D)$ denote the rate-distortion function for the source $X$. See \cite{ShannonRD} for a definition.  Shannon  \cite{ShannonRD} uses an expectation condition when defining the rate-distortion function
\begin{align}\label{EED}
\lim_{n \to \infty} E \left [ \frac{1}{n} d^n(X^n, Y^n)  \right ]  \leq D
\end{align}
The definition that we use for distortion is the limit of (\ref{PED}) as block-length $n \to \infty$, that is,
\begin{align}\label{PEDLimit}
\lim_{n \to \infty} \Pr \left ( \frac{1}{n} d^n(X^n, Y^n) > D \right ) = 0 
\end{align}
These two rate-distortion functions are the same as proved in \cite{AgarwalSahaiMitter}. The dependence of the rate-distortion function on the distortion metric is not shown explicitly.

\emph{Notation: }The rate-distortion function for source $X'$ with distortion $D'$ will be denoted by $R_{X'}(D')$.

\begin{thm}\label{MX}
Given a system where i.i.d. source $X^n$ is communicated to within a distortion $D$ under metric $d$ to with error probability $< \epsilon$.  Let  $R = R_{X}(D) - \alpha$ for some $\alpha > 0$. Then, rate $R$ source $M^n$ (where $M^n$ is arbitrary) can be communicated under the MBP criterion with error probability $< \delta_n$ from user $s$ to user $r$, \emph{in place of} communicating source $X^n$  by using the method described in Section \ref{Spirit}, for some $\delta_n \to \epsilon$ as $n \to \infty$. 
\end{thm}
\begin{pf}
This follows from Theorem 1 in \cite{AgarwalSahaiMitter}. Note that the codes are generated i.i.d. $X$ and hence, the distribution of $X^n$ is mantained as required in Section \ref{Spirit}. Also note that Theorem 1 in \cite{AgarwalSahaiMitter} is universal: the channel might be unknown. Thus, for this theorem, the medium might be unknown, as required in Section \ref{Spirit}. 
\end{pf}

%\begin{center}
%\includegraphics[scale = 1.0]{MXCut.pdf}
%\end{center}

We use the above theorem to prove the result concerning communication of source $X'$ of block length $n'$ to within a distortion $D'$ under metric $d'$ with error $< \epsilon'_n$ from user $s$ to user $r$ \emph {in place of} i.i.d. $X$ source of block length $n$ which is known to be communicated to within a distortion $D$ under metric $d$ with error probability $< \epsilon$ from user $s$ to user $r$.

\begin{thm}\label{X'X}
Given that  i.i.d. $X$ source of block length $n$ is communicated from user $s$ to user $r$ to within a distortion $D$ under metric $d$ with error probability $< \epsilon$. Let $\displaystyle{\frac{n}{n'} > \frac{R_{X'}(D')}{R_{X}(D)} + \psi}$ for some $\psi > 0$. Then  i.i.d. $X'$ source of block length $n'$ can be communicated from user $s$ to user $r$  to within a distortion $D'$ under metric $d'$ with error probability $<\epsilon'_n$ for some $\epsilon'_{n} \to \epsilon$ as $n \to \infty$ \emph{in place of i.i.d. $X$ source}. 
\end{thm}
\begin{pf}
%\begin{center}
%\includegraphics[scale = 0.9]{XdashXCut.pdf}
%\end{center}
This uses the usual argument of source-coding followed by reliable channel coding. Roughly, the argument is the following. Compress the source $X'$ to within the distortion level $D$. The output is a message set of cardinality $2^{mR_{X'}(D')}$. Communicate the compressed message over the system from user $s$ to user $r$. The message gets communicated correctly with probability $1 - \epsilon$.  This communication with probability $1 - \epsilon$ can be accomplished because the conditions of the previous theorem, Theorem \ref{MX} are satisfied.  Finally, decode the source. End to end, the required communication of i.i.d. $X'$ source is accomplished.  More precisely, there exist  
\newline
source encoder $\displaystyle{s^{n'}_{e}: \mathcal X^{n'} \rightarrow \mathcal M^{n'} = \{1, 2, \ldots 2^{n' (R_{X'}(D') + \frac{\psi}{2})} \}  }$ and
\newline
source decoder $\displaystyle{s^{n'}_{d}: \mathcal M^{n'} = \{1, 2, \ldots 2^{n' (R_{X'}(D') + \frac{\psi}{2})} \}  \rightarrow \mathcal Y^{n'}}$ such that 
\begin{align}
\Pr \left ( \frac{1}{n} d^n(X'^{n'}, s^{n'}_d \circ s^{n'}_e (X'^{n'}) > D' \right ) = \eta_{n'} \to 0   \ \mbox{as} \ n' \to \infty
\end{align}
By an assumption in the theorem, it follows that 
\begin{align}
nR_X(D) - n' \left [ R_{X'}(D') + \frac{\psi}{2} \right ] > n \left [  \frac{R_X(D)}{R_X(D) + \psi} \frac{\psi}{2}  \right ]
\end{align}
Define $\displaystyle{ \alpha =   \left [  \frac{R_X(D)}{R_X(D) + \psi} \frac{\psi}{2}  \right ] }$. It follows that 
\begin{align}
n(R_X(D) - \alpha) > n' \left [ R_{X'}(D') + \frac{\psi}{2} \right ]
\end{align}
We can, thus, think of the maps $s^{n'}_e$ and $s^{n'}_d$, as
source encoder $\displaystyle{s^{n'}_{e}: \mathcal X^{n'} \rightarrow \mathcal M^{n} = \{1, 2, \ldots 2^{n (R_{X}(D)  - \alpha) } \}  }$ and
\newline
source decoder $\displaystyle{s^{n'}_{d}: \mathcal M^{n} = \{1, 2, \ldots 2^{n' (R_{X'}(D') - \alpha ) } \}  \rightarrow \mathcal Y^{n'}}$. This is because $M^{n'} \subset M^n$. We can then re-label, and call $s^{n'}_{e}$ as $s^n_e$, and call $s^{n'}_d$ as $s^n_d$.

First compress the source $X'^{n'}$ using the source encoder $s^{n}_e$. The output $M^n = s^{n}_e (X'^{n'})$ is some  distribution on $\mathcal M^n$.
By Theorem \ref{MX}, it follows that there exists encoder $c_e^n$ and decoder $c_d^n$ such that with these encoder and decoder,  $M^n$ of rate $R_{X}(D) - \alpha$ is communicated under the $MBP$ criterion with error $< \xi_n$. from user $s$ to user $r$
 where $\xi_n \to \epsilon$ as $n \to \infty$. The decoding of $M^n$ at user $r$ is $\hat M ^n$. Now apply the source-decoder $s_d^n$ to $\hat M^n$. We get a decoding $Y'^{n'}$ of source $X'^{n'}$. End to end, 
\begin{align}
\Pr \left ( \frac{1}{n} d^n(X'^{n'}, Y'^{n'}) > D'  \right ) \leq \xi_n  +  \eta_n  = \epsilon'_{n}   \to \epsilon \ \mbox{as} \ n \to \infty
\end{align}
This proves the theorem.

\end{pf}

Now, we specialize this theorem to the case when $X'$ has the same distribution as $X$ and is independent of $X$. 

\begin{thm}\label{XX}
Given that  i.i.d. $X$ source of block length $n$ is communicated over the system  to within a distortion $D$ under metric $d$ with error $< \epsilon$  from user $s$ to user $r$. Let $D' < D$ and $R_{X}(D') < R_X(D)$ (note: strictly less). Then, i.i.d. $X$ source of block length $n$ can be communicated over the system to within a distortion $D'$ under metric $d$ with error probability $< \epsilon'_n$ from user $s$ to user $r$ by using an architecture which consists of source compression of $X$ followed by communication of the compressed source under the MBP criterion with some error probability. By use of this new architecture, end-to-end, the i.i.d. $X$ source is communicated to within a distortion level $D'$ under the metric $d'$ with error $\epsilon_n \to \epsilon$ as $n \to \infty$. The communication of sources from user $i$ to user $j$, $(i,j) \neq (s,r)$ is not affected by the new architecture. That is, for $(i,j) \neq (s,r)$, if $X_{ij}$ is received as $Y_{ij}$ in the given architecture, it is received precisely as $Y_{ij}$ in the new architecture also. The energy and bandwidth consumption in the two architectures is the same. 
\end{thm}
\begin{pf}
%\begin{center}
%\includegraphics[scale = 0.9]{XXCut.pdf}
%\end{center}
This can be proved by use of the previous theorem, Theorem \ref{X'X}  with $X' = X$, $n' = n$ and $D' = D'$ as follows. 

$\displaystyle{\frac{n}{n'} = \frac{n}{n} = 1  > \frac{R_{X}(D')}{R_{X}(D)}  + \frac{1}{2} \frac{R_X(D) - R_X(D')}{{R_X(D)}}}$.

Theorem \ref{X'X} applies with $\psi =  \frac{1}{2} \frac{R_X(D) - R_X(D')}{{R_X(D)}}$.

There exist $s_e^n, s_d^n, c_e^n, c_d^n$ as in the  previous theorem. The new architecture consists of modem $h'_s = h_s \circ c_e^n \circ s_e^n$ at user $s$ and modem $h'_r = s_d^n \circ c_d^n \circ h_r$ at user $r$. The required communication of source $X^n$ from user $s$ to user $r$ in the new architecture occurs by using modem $h'_s$ and $h'_r$ at users $s$ and $r$. Modems for rest of the users remain unchanged. $h'_s$ can be interpreted as follows. First, source $X^n$ is compressed using $s_e^n$. The compressed source $M^n$ is encoded by use of $h_s \circ c_e^n$ so that it is communicated to with  maximal block error probability $ \leq  \epsilon)$. $h'_r$ can be interpreted as follows. First the received sequence $Y^{s,n}$ is decoded  into $\hat M^n$ by use of $c_d^n \circ h_r$. $\hat M^n$ is the estimate of $M^n$ with maximal block error probability  $\leq \epsilon)$. Then, $\hat M^n$ is source-decoded using $s_d^n$. End-to-end, the source $X^n$ is communicated to within distortion level $D'$ under the metric $d$ with error probability $<\epsilon'_n$. from user $s$ to user $r$ such that $\epsilon'_n \to \epsilon$ as $n \to \infty$.

The rest of the statements in the theorem follow from the discussion in Section \ref{Spirit}. This completes the proof.
\end{pf}

\emph{Note:} The total time incurred in the end-to-end communication of i.i.d. $X$ source (the delay)  might be larger in the separation architecture as compared to the original architecture. However, this does not concern us.

\emph{Note:} We have provided a separation architecture for communication of source $X_{sr}$ to within a distortion level $D$ under metric $d$ but it is built on top of the existing architecture which already accomplishes precisely the same communication!  This is just a proof technique.  The proof follows a route of ``building on top'' of the existing architecture. This helps to prove that there is no loss of optimality in using separation architectures. In practice one can use other separation architectures which are not ``building on top'' of the existing architecture.

\section{Application to Information Theory: Separation for rate-distortion in networks} \label{InformationTheory}

In this Section, we prove a source-channel separation theorem for rate-distortion for networks. Subsection \ref{Discussion} contains a discussion with examples. 
\subsection{Universal source-channel separation for rate-distortion in networks}

Information theory is concerned with the behavior of  quantities in the limit as the block-length $n \to \infty$. As stated before, delays do not matter.

We first consider the question of communication of i.i.d. $X$ source $X^n$ from user $s$ to user $r$  to within a distortion level $D$ as block-length $n \to \infty$. The modem at user $i$ is $h_i^n, 1 \leq i \leq N$, when the block-length is $n$. We make statements concerning optimal architectures for  this communication.  It is required, as stated in Section \ref{Spirit} that the communication between other users is not affected.

\emph{Definition:}
When the block-length is $n$, modem $h_i^n$ is used at user $n$, $ 1 \leq i \leq N$. The input that needs to be communicated from user $s$ to user $r$ is $X^n$. The reproduction of $X^n$ at user $r$ is $Y^n$. We say that i.i.d. $X$ source is communicated to within a distortion level $D$ over the system from user $s$ to user $r$ if (\ref{PEDLimit}) holds.

\begin{thm} \label{Separation}
Let there exist modems $h_i^n, 1 \leq i \leq N, 1 \leq n \leq \infty$ such that i.i.d. $X$ source is communicated from user $s$ to user $r$ to within a distortion level $D$. Let $D'>D$ be such that $R_X(D') < R_X(D)$. Then, there exist modems ${h'}_i^n, 1 \leq i \leq N, 1 \leq n \leq \infty$ such that modem ${h'}_i^n$ at user $i$, 
\begin{enumerate}
\item
${h'}_s^n$ first source-codes i.i.d. $X$ source $X^n$ of block length $n$ and this is followed by reliable communication of the resulting message to user $r$. 
\item 
${h'}_r^n$ does channel decoding followed by source decoding to get a decoding $Y^n$ of $X^n$. 
\item
${h'}_i^n$ consumes the same energy and bandwidth as $h_i^n$ for all $i$, for all $n$.
\item
communication of sources between other users in not affected in the sense defined in Section \ref{Spirit}: $X_{ij}$ is received precisely as $Y_{ij}$ for $(i,j) \neq (s,r)$ even if modemss ${h'}_i^n$ are used instead of $h_i^n, 1 \leq i \leq N$ 
\end{enumerate}
\end{thm}
\begin{pf}
This follows immediately from Theorem \ref{XX}. 
\end{pf}

Now, we prove a network version of the above theorem: communication to particular distortion levels is desired between various users, not just from user $s$ to user $r$.

\emph{Notation and definitions:} Let $\mathcal A \subset \{ (i,j)\ | \ 1 \leq i,j \leq N, i \neq j\}$. Let $(p,q) \in \mathcal A$. Let $d_{pq}: \mathcal X_{pq} \times \mathcal Y_{pq} \rightarrow [0,\infty)$ be a distortion metric as in Section \ref{Distortion}. $d^n_{pq}$ is additive average distortion defined in the same way as $d^n_{sr}$ is defined in Section \ref{Distortion}. Communication of source $X_{pq}$ to within a distortion level $D_{pq}$ under distortion metric $d_{pq}$ is defined analogously to (\ref{PEDLimit}).

\begin{thm}\label{SeparationNetwork}
Let there exist modems $h_i^n, 1 \leq i \leq N, 1 \leq n \leq \infty$ such that for all $(p,q) \in \mathcal A$, i.i.d. source $X_{pq}$ is communicated from user $p$ to user $q$ to within a distortion level $D_{pq}$. Let $D'_{pq} >D_{pq}$ be such that $R_{X_{pq}}(D'_{pq}) < R_{X_{pq}}(D_{pq}) \forall p,q \in \mathcal A$. Then, there exist modems ${h'}_i^n, 1 \leq i \leq N, 1 \leq n \leq \infty$ such that  modems ${h'}_i^n$ at user $i$ satisfy the following:
\begin{enumerate}
\item
${h'}_p^n$ first source-codes i.i.d. $X_{pq}$ source $X^n_{pq}$ of block length $n$ and this is followed by reliable communication of the resulting message to user $q$. 
\item 
${h'}_q^n$ does channel decoding followed by source decoding to get a decoding $Y^n_{pq}$ of $X^n_{pq}$. 
\item
${h'}_i^n$ consumes the same energy and bandwidth as $h_i^n$ for all $i$, for all $n$.
\item
communication of sources between other users in not affected in the sense defined in Section \ref{Spirit}. 
\end{enumerate}
\end{thm}

\begin{pf}
This can be done step by step. First carry out the separation procedure for one user pair $(p_1, q_1)$ in $\mathcal A$. This can be done by the previous theorem, Theorem \ref{Separation}. After making this change of architecture, source $X_{p_1q_1}$ is still being communicated to within a distortion level $D_{p_1q_1}$ from user $p_1$ to user $q_1$. Very important, is the fact that sources $X_{ij}, (i,j) \neq (p_1, q_1)$ are still being received as $Y_{ij}$. In particular, for $(p,q) \in \mathcal A \smallsetminus (p_1, q_1)$, $X_{pq}$ is still communicated to within a distortion level $D_{pq}$ over the system. Now choose another user pair $(p_2, q_2) \in \mathcal A \smallsetminus (p_1, q_1)$ and repeat the procedure until all user pairs in $\mathcal A$ are exhausted. This completes the proof.
\end{pf}

%Note: The sources $X_{pq}$ need not have the same block-length. Source $X_{pq}$ might have block length $n_{pq}$. We have avoided making a completely general statement to prevent complexity in notation. However, the statement is still a statement about communication in the limit that all block-lengths $n_{pq} \to \infty$. 

A high-level version of this theorem is the source-channel separation theorem for rate-distortion for networks when the sources that various users want to communicate to each other are independent of each other. 

\begin{thm}\label{HighLevelSeparationNetwork}
Consider a medium $m$ and $N$ users. $N$ might change with time. \emph{Independent} sources $X_{ij}$ are communicated from user $i$ to user $j$, $1 \leq i,j \leq N$, $i \neq j$, over the medium. $X_{ij}$ is transmitted at user $i$ and received at user $j$. Let $\mathcal A$ be a subset of user pairs, that is, $\mathcal A \subset \{ (i,j)\ | \ 1 \leq i,j \leq N, i \neq j\}$. For $(p,q) \in \mathcal A$, it is known that $X_{pq}$ is i.i.d. It is required to communicate sources $X_{pq}, (p,q) \in \mathcal A$ to within a distortion level $D_{pq}$ over the system under a distortion metric $d_{pq}$. In order to accomplish this communication, it is sufficient to consider separation architectures:  that is, architectures which compress i.i.d. source $X_{pq}$ , $(p,q) \in \mathcal A$ to within the desired distortion level and then communicate the compressed message reliably over the system. Communication of other sources is not affected in the separation architecture in the sense that if $X_{ij}$, $(i,j) \notin \mathcal A$, and if $X_{ij}$ is received as $Y_{ij}$ in the original archicture,  $X_{ij}$ is received precisely as $Y_{ij}$ in the separation architecture too. Of course, $X_{pq}, (p,q) \in \mathcal A$ is not necessarily received as $Y_{pq}$ in the separation architecture. However, it is received as some $Y'_{pq}$ which is to within a distortion $D_{pq}$ of $X_{pq}$. Energy and bandwidth consumption remains the same at each user. Delay incurred for communication of sources $X_{ij}, (i,j) \notin \mathcal A$ remains the same.
\end{thm}

\section{Discussion and Conclusion} \label{Discussion}

We have proved a source-channel separation theorem for rate-distortion in the network setting when the sources that various users wish to communicate with  each other are independent of each other.  Note that the medium is unknown. Assuming that random-coding is permitted, for every encoding-decoding scheme which achieves the required distortion bounds over the medium, we have demonstrated the existence of a separation based scheme which has the same performance as the original scheme, and this does not require the knowledge of the medium. What the result says, then, is that for the problem of rate-distortion communication over an unknown medium, it is sufficient to restrict attention to separation based protocols.

For example, consider the case of the internet. Different users wish to communicate various sources to each other. Different sources have different distortion requirements. For example, one user might want to communicate an e-mail to another user, for which no distortion is allowed. Another user might be chatting via voice or via video with another user, and in that case, distortion is permitted. The distortion metric in the case of voice and video is not additive, but for sake of the argument, suppose that that was the case. The structure of the internet is unknown. In fact, it changes with time. We still need to design a protocol to meet the desired communication requirements. What we prove is that if random-coding is permitted and sources that different users want to communicate to other users are independent of each other, it is sufficient to restrict attention to separation based protocols. 

Another example is wireless communication. Wireless medium is time varying  and unknown. Users want to communicate voice which allows distortion. For sake of the argument, assume that the distortion metric for voice is additive. There exist various protocols for wireless communication, for example, CDMA and GSM. It is a reasonable assumption that what different users talk is independent of each other. We prove that assuming that random-coding is permitted, one does not lose anything by restricting attention to separation-based protocols for the question of the number of users which can be communicating over the wireless medium at a particular time.

The above problem of communicating sources with a fidelity criterion when the sources are not independent is open in general. Source-channel separation based architectures are not optimal in general.

\bibliography{TR}

\end{document}